\documentclass[prb,letter,twocolumn,floatfix,superscriptaddress]{revtex4-2}
\usepackage{epsfig}
\usepackage{amsmath,amssymb}
\usepackage{graphicx}
\usepackage[dvipsnames,usenames]{xcolor}
\usepackage[normalem]{ulem}
\usepackage{soul}  
\usepackage{hyperref}
\hypersetup{colorlinks,bookmarksopen,bookmarksnumbered,
citecolor=red,
linkcolor=red,
pdfstartview=false,
urlcolor=red}
\emergencystretch=\maxdimen
\begin{document}


\title{Interactions and integrability in weakly monitored Hamiltonian systems} 



\author{Bo Xing}
\email{bo\_xing@mymail.sutd.edu.sg}
\affiliation{Science, Mathematics and Technology Cluster, Singapore University of Technology and Design, 8 Somapah Road, 487372 Singapore} 
\author{Xhek Turkeshi} 
\affiliation{JEIP, UAR 3573 CNRS, Coll\`ege de France, PSL Research University, 11 Place Marcelin Berthelot, 75321 Paris, France}
\affiliation{Institut f\"ur Theoretische Physik, Universit\"at zu K\"oln, Z\"ulpicher Strasse 77, 50937 K\"oln, Germany} 
\author{Marco Schir\'{o}} 
\affiliation{JEIP, UAR 3573 CNRS, Coll\`ege de France, PSL Research University, 11 Place Marcelin Berthelot, 75321 Paris, France}
\author{Rosario Fazio}
\affiliation{The Abdus Salam International Centre for Theoretical Physics (ICTP), Strada Costiera 11, 34151 Trieste, Italy}
\affiliation{Dipartimento di Fisica, Universit\`{a} di Napoli “Federico II”, Monte S. Angelo, I-80126 Napoli, Italy}
\author{Dario Poletti} 
\email{dario\_poletti@sutd.edu.sg}
\affiliation{Science, Mathematics and Technology Cluster, Singapore University of Technology and Design, 8 Somapah Road, 487372 Singapore}
\affiliation{Engineering Product Development Pillar, Singapore University of Technology and Design, 8 Somapah Road, 487372 Singapore}
\affiliation{Centre for Quantum Technologies, National University of Singapore 117543, Singapore}

\begin{abstract} 
Interspersing unitary dynamics with local measurements results in measurement-induced phases and transitions in many-body quantum systems. When the evolution is driven by a local Hamiltonian, two types of transitions have been observed, characterized by an abrupt change in the system size scaling of entanglement entropy. The critical point separates the strongly monitored area-law phase from a volume law or a sub-extensive, typically logarithmic-like one at low measurement rates. 
Identifying the key ingredients responsible for the entanglement scaling in the weakly monitored phase is the key purpose of this work.
For this purpose, we consider prototypical one-dimensional spin chains with local monitoring featuring the presence/absence of U(1) symmetry, integrability, and interactions.
Using exact numerical methods, the system sizes studied reveal that the presence of interaction is always correlated to a volume-law weakly monitored phase. In contrast, non-interacting systems present sub-extensive scaling of entanglement. Other characteristics, namely integrability or U(1) symmetry, do not play a role in the character of the entanglement phase.     
\end{abstract}

\maketitle

\textit{Introduction.---} 
Monitoring the otherwise unitary evolution of a many-body quantum system substantially changes its dynamical features~\cite{skinner2019measurementinducedphase,li2018quantumzenoeffect,li2019measurementdrivenentanglement,cao2019entanglementina,szyniszewski2019entanglementtransitionfrom,szyniszewski2020universalityofentanglement}. 
Indeed, quantum measurements are inherently stochastic and non-linear, as they collapse the system's state onto the outcome eigenspaces, thus inducing a non-equilibrium behavior that is non-unitary and described by quantum trajectories~\cite{breuer2002thetheoryof,wiseman2009quantummeasurementand,jacobs2014quantummeasurementtheory}. 
When monitoring local variables, measurements may disentangle degrees of freedom, thus competing with the entangling unitary evolution. 
This tension results in the hallmark phenomenon of measurement-induced phase transitions (MIPT) in the structure of the typical quantum trajectory, cf. Ref. ~\cite{fisher2023randomquantumcircuits,potter2022quantumsciencesandtechnology,lunt2022quantumsciencesandtechnology} for reviews. 
Crucially, the features of the typical trajectory are generally invisible in the average dynamics, described by a master equation or quantum channel, and revealable by non-linear functionals of the quantum state. 
For instance, the system size scaling of the entanglement entropy has been extensively used to characterize the various measurement-induced phases~\cite{bao2020theoryofthe,choi2020quantumerrorcorrection,gullans2020dynamicalpurificationphase,gullans2020scalableprobesof,bao2021symmetryenrichedphases,hoke2023quantuminformationphases,koh2022experimentalrealizationof,noel2021measurementinducedquantum}. The strongly monitored phase generally presents an area-law scaling, while the weakly monitored one exhibits either (extensive) volume law or a sub-extensive, typically logarithmic-like scaling. 
More recently, in \cite{tirrito2023fullcountingstatistics} it was shown that one can use other quantities, like second-order cumulants, to locate the transition point.  

\begin{figure}[h!]
    \includegraphics[width=\columnwidth]{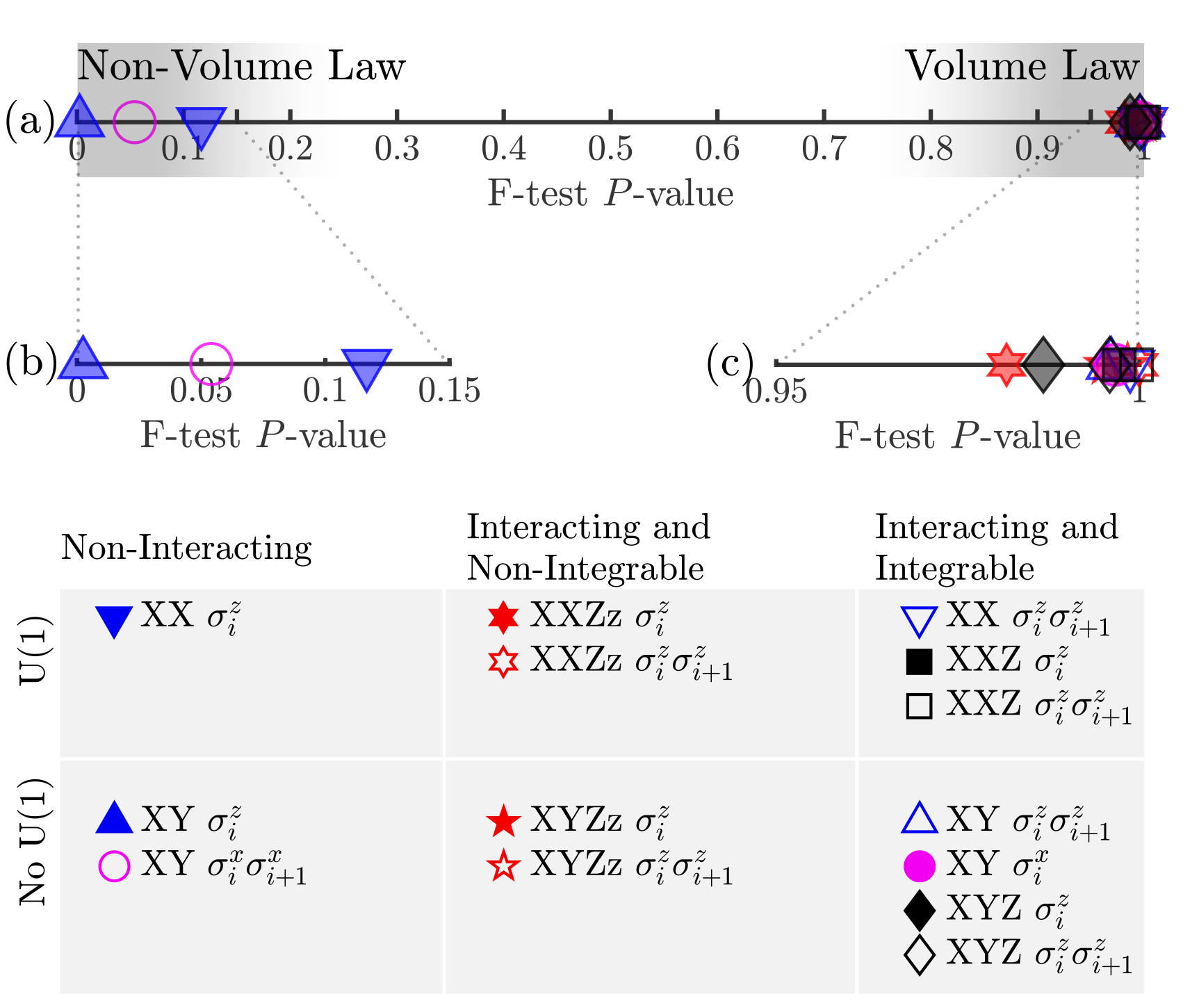}
    \caption{(a) F-test $P$-value of the models considered in this work, grouped in the table below.
    A value of F-test $P \rightarrow 1$ gives a high confidence that the underlying statistical model grows linearly with $L$ (volume law).
    Conversely, $P \rightarrow 0$ suggests that the fit is more likely to be linear in $\ln{L}$ (logarithmic law).
    The regions corresponding to non-volume law and volume law are enlarged in (b) and (c) respectively. 
    }
    \label{fig:F-test}
\end{figure}

At the same time, different unitary dynamics can lead to different scalings of the entanglement.  
Random unitary circuits have been extensively investigated as a useful toy model to understand key aspect of MIPTs~\cite{li2021entanglementdomainwalls,zabalo2022operatorscalingdimensions,sierant2022universalbehaviorbeyond,oshima2023chargefluctuationand,agrawal2022entanglementandcharge,chiriaco2023diagrammaticmethodfor,turkeshi2020measurementinducedcriticality,weinstein2022scramblingtransitionin,kelly2022coherencerequirementsfor,block2022measurementinducedtransition,sharma2022measurementinducedcriticality,vijay2020measurementdrivenphase,fan2021selforganizederror,ippoliti2022fractallogarithmicand,nahum2021measurementandentanglement,zhu2023structured}.
Their random and structureless nature allowed for essential breakthrough using replica methods~\cite{skinner2019measurementinducedphase,nahum2021measurementandentanglement,jian2020measurementinducedcriticality,li2021statisticalmechanicsmodel,bao2020theoryofthe} backed up by large-scale numerical simulations~\cite{zabalo202criticalpropertiesof,turkeshi2020measurementinducedcriticality,lunt2021measurementinducedcriticality,sierant2022measurementinducedphase}. 
Nevertheless, archetypal many-body quantum systems evolve through a Hamiltonian, whose conservation laws and structure constrain the explored Hilbert space. 
For instance, without measurement, the stationary state of interacting systems locally thermalizes when only energy is conserved~\cite{Srednicki1999,Rigol_2008,silvareview,foini,pappalardi2023general,Pappalardi_2022}, and relaxes to a gran-canonical or generalized Gibbs ensemble when  more (possibly infinite) conservation laws are present~\cite{Essler_2016}. 

A key open question is whether monitored integrable Hamiltonians, even when interacting, have a sub-extensive or extensive scaling of entanglement entropy at low measurement strength \cite{fuji2020measurementinducedquantum}. 
Currently, a phase transition between volume and area-law entanglement scaling has been observed in  Ref.~\cite{fuji2020measurementinducedquantum,tang2020measurementinducedphase,lunt2020measurementinducedentanglement}, while Ref.~\cite{cao2019entanglementina,fidkowski2021howdynamicalquantum,coppola2022growthofentanglement,doggen2023evolutionofmany,dogger2022generalizedquantummeasurements} showed that non-interacting models cannot sustain a stable volume law phase if no-postselection is present~\cite{gal2023volumetoarea,granet2022volumelawtoarealaw}, with the weakly monitored phase having a sub-extensive entanglement scaling fixed by the system's symmetries~\cite{nahum2020entanglementanddynamics,fava2023nonlinearsigmamodels,jian2022criticalityandentanglement,jian2023measurementinducedentanglement,poboiko2023theoryoffree,buchhold2021effectivetheoryfor,loio2023purificationtimescalesin,merritt2023entanglementtransitionswith,klocke2023majorana,alberton2021entanglementtransitionin,turkeshi2021measurementinducedentanglement,zhou2021nonunitaryentanglementdynamics,minoguchi2022continuousgaussianmeasurements,botzung2021engineereddissipationinduced,tsitsishvili2023measurement}.
However, despite these early results, a clear picture of the interplay between interaction, integrability and symmetry in monitored dynamics is missing. 

This work elaborates in this direction by investigating the measurement-induced evolution of archetypal one-dimensional spin chains.
Varying the parameters, we study all combinations of different scenarios: whether they are integrable, interacting, have a U(1) symmetry, or not. Given the computational complexity of these models, we use an exact diagonalization approach and study up to $L = 28$ spins. For the models and sizes studied, we observe that all models with interactions, independent of whether they have a U(1) symmetry or they are integrable, manifest a volume law scaling weakly monitored phase. On the other hand, all non-interacting models have a sub-extensive (logarithmic) scaling of entanglement entropy.

\textit{Model and method.---}
We study a generalized 1D Heisenberg spin chain of length $L$ with open boundary conditions and Hamiltonian~\footnote{We denote by $\hat{\sigma}^\alpha_i$ the $\alpha=x,y,z$ Pauli matrix acting on the $i$-th spin. }
\begin{align}\label{eqn:hamiltonian}
    \hat{H} = &\sum_{l=1}^L \left[\left(\sum_{\alpha={x,y,z}}J_{\alpha} \hat{\sigma}^{\alpha}_{l}\hat{\sigma}^{\alpha}_{l+1}\right) + h_{z} (-1)^{l} \hat{\sigma}^{z}_{l}\right] \;.  
\end{align}
This Hamiltonian can become interacting (non-interacting), integrable (non-integrable), or U(1)-conserving (non-conserving) by simply varying the values of the couplings $J_\alpha$ and of the staggered field $h_z$.
With $J_z = 0$ the Hamiltonian is non-interacting~\cite{Franchini_2017,eckle2019models,Fradkin2013}, it is interacting and integrable when $h_z=0$, and it conserves the total magnetization $\hat{S}_z=\sum_{i=1}^L \hat{\sigma}_i^z$ when $J_x=J_y$.
Following the literature, we refer to these models with different names: XX when $J_x=J_y$ and $h_z=J_z=0$, XY for $J_x \neq J_y$ and $h_z=J_z=0$, XXZ for $J_x=J_y$, $J_z \neq 0$ and $h_z=0$, XYZ for $J_x \neq J_y \neq J_z \neq 0$, and $h_z=0$, XXZz for $J_x = J_y$, $J_z \neq 0$ and $h_z \neq 0$ and XYZz for $J_x \neq J_y\neq J_z \neq 0$ and $h_z \neq 0$. 

We consider the system coupled to independent homodyne detectors on the local operators $\hat{O}_l$ such that $\hat{O}_l = \hat{O}_l^\dagger$ and $\hat{O}_l^2 = \openone$. 
The resulting quantum evolution is given by the stochastic Schr\"odinger equation~\cite{wiseman2009quantummeasurementand,jacobs2014quantummeasurementtheory}
\begin{equation}
\begin{split}\label{eqn:propagator}
    d|\Psi_t\rangle &= -i \hat{H} dt |\Psi_t\rangle  - \frac{1}{2}\gamma dt\sum_{l} \left(\hat{O}_l-\langle \hat{O}_l\rangle_t\right)^2 |\Psi_t\rangle \\  &+ \sum_{l} d\xi_l(t) \left(\hat{O}_l-\langle \hat{O}_l\rangle_t\right)|\Psi_t\rangle
   \;,
\end{split}
\end{equation}
where $\gamma$ is the measurement rate and $\langle O \rangle_t\equiv \langle \Psi_t |\hat{O}_l |\Psi_t\rangle $ is the expectation value on the trajectory $|\Psi_t\rangle\equiv |\Psi_t(\xi)\rangle $ at time $t$, and the $d\xi_l$ are \^Ito Gaussian noise with average $\overline{d\xi_l} = 0$ and $d\xi_l(t) d\xi_{l'}(t') = \delta_{l,l'} \delta_{t,t'} \gamma dt.$
In this work, we study the monitoring of local or nearest-neighboring operators. Specifically we choose $\hat{O}_l = \hat{\sigma}_l^\alpha$ with $l=1,\dots,L$ or $\hat{\sigma}_l^\alpha \hat{\sigma}_{l+1}^\alpha$ with $l=1,\dots,L-1$. 
The choice of operators for the various Hamiltonian models is summarized in Fig.~\ref{fig:F-test}. 
It is important to stress that non-interacting Hamiltonian may lead to interacting (non-Gaussian) dynamics depending on the measured operators. 
For instance, interspersing the XX unitary evolution with $\hat{\sigma}_{l}^{z}\hat{\sigma}_{l+1}^{z}$ measurements leads to interaction effects. We also note that monitoring the $\hat{\sigma}_l^x$ during the XY dynamics also leads to interactions, as it can be understood via the Jordang-Wigner transformation ($\hat{\sigma}_l^x$ is a string of fermions). Instead, monitoring $\hat{\sigma}_{l}^{x}\hat{\sigma}_{l+1}^{x}$ preserve the non-interacting cahracter of the XY model. In all our investigations we use a monitoring rate $\gamma=0.1$, for which a behavior different from area law can be observed. At even smaller $\gamma$, the non volume-law phase would appear as a volume law for the system sizes analyzed.
On the other hand, for higher $\gamma$, the measurements become too strong to allow the volume-law phase to emerge.   

Starting from an initial random product state, we evolve the wavefunction using Eq.~\ref{eqn:propagator}, for systems reaching up to $L=28$ spins.
In our implementation, we exponentiate Eq.~\eqref{eqn:propagator} using \^Ito calculus and trotterize up to second order in $dt$~\cite{wiseman2009quantummeasurementand}. The resulting implementation is, therefore, a quantum circuit, where each layer is composed of a unitary step and a measurement step. 
At the time $t=T$ 
we compute the half-chain entanglement entropy for each trajectory~\cite{amicorev,Calabrese2004}
\begin{align}\label{eq:defent}
    S_\xi(L/2) \equiv  -\mathrm{tr}\left( \hat{\rho}^\mathrm{sub}_t \ln \hat{\rho}^\mathrm{sub}_t\right), 
\end{align}
where $\hat{\rho}^\mathrm{sub}_t= \mathrm{tr}_{L/2+1,L}( |\Psi_t\rangle \langle \Psi_t|)$ is the reduce density matrix obtained tracing out spins from $L/2+1$ to $L$. 
Being a fluctuating quantity, we study the mean $S(L/2)\equiv \overline{S_\xi}$, with the average being over the $N_\mathrm{tj}$ initial random product states, the corresponding $N_\mathrm{tj}$ evolutions ($N_\mathrm{tj}\geq 160$ for $L\le 24$ and $N_\mathrm{tj}\geq 80$ for $L>24$), and over the $N_T=5$ chosen values of $T = (25+k)/J_x$ with $k=1,\dots,N_T$. 
(We have preliminarily checked that at the time $T=25/J_x$ the average entanglement entropy is converged to its stationary value. 
If $S(L/2)$ grows linearly with $L$ ($\ln(L)$), the entanglement entropy obeys the volume law (logarithmic law).

We corroborate the scaling of $S(L/2)$ in a systematic fashion by using the $F$-test~\cite{Mood1950} to quantitatively determine the statistical likelihood of a model presenting a volume law phase or not. 
The $F$-test is a likelihood-ratio test that assesses the goodness of fit of two competing statistical models.
This likelihood ratio is obtained by evaluating a test statistic equation that follows the $F$-distribution.
For each set of $S(L/2)$ of a model, we produce two curves $\tilde{S}_{L}$ and $\tilde{S}_{\ln(L)}$, which are linear best-fits curves of $S(L/2)$ in $L$ and $\ln(L)$ respectively.
These best-fit lines help us compute the $F$-test statistic,
\begin{equation}
    F = \frac{\tilde{E}_{L}}{\tilde{E}_{\ln(L)}}
,\end{equation}
where $\tilde{E}_{L}$ ($\tilde{E}_{\ln(L)}$) is the sum of the squared errors between $S(L/2)$ and $\tilde{S}_{L}$ ($\tilde{S}_{\ln(L)}$).
Since we expect the most likely statistical model to have a lower $\tilde{E}$, a lower $F$ will favor the null hypothesis (volume law), and a higher $F$ will favor the new hypothesis (logarithmic/area law).
The probability of finding a test statistic as high, or higher than $F$ is thus referred to as the $P$-value.
The smaller the $P$-value, the more significant the new hypothesis becomes.

\begin{figure}
    \includegraphics[width=\columnwidth]{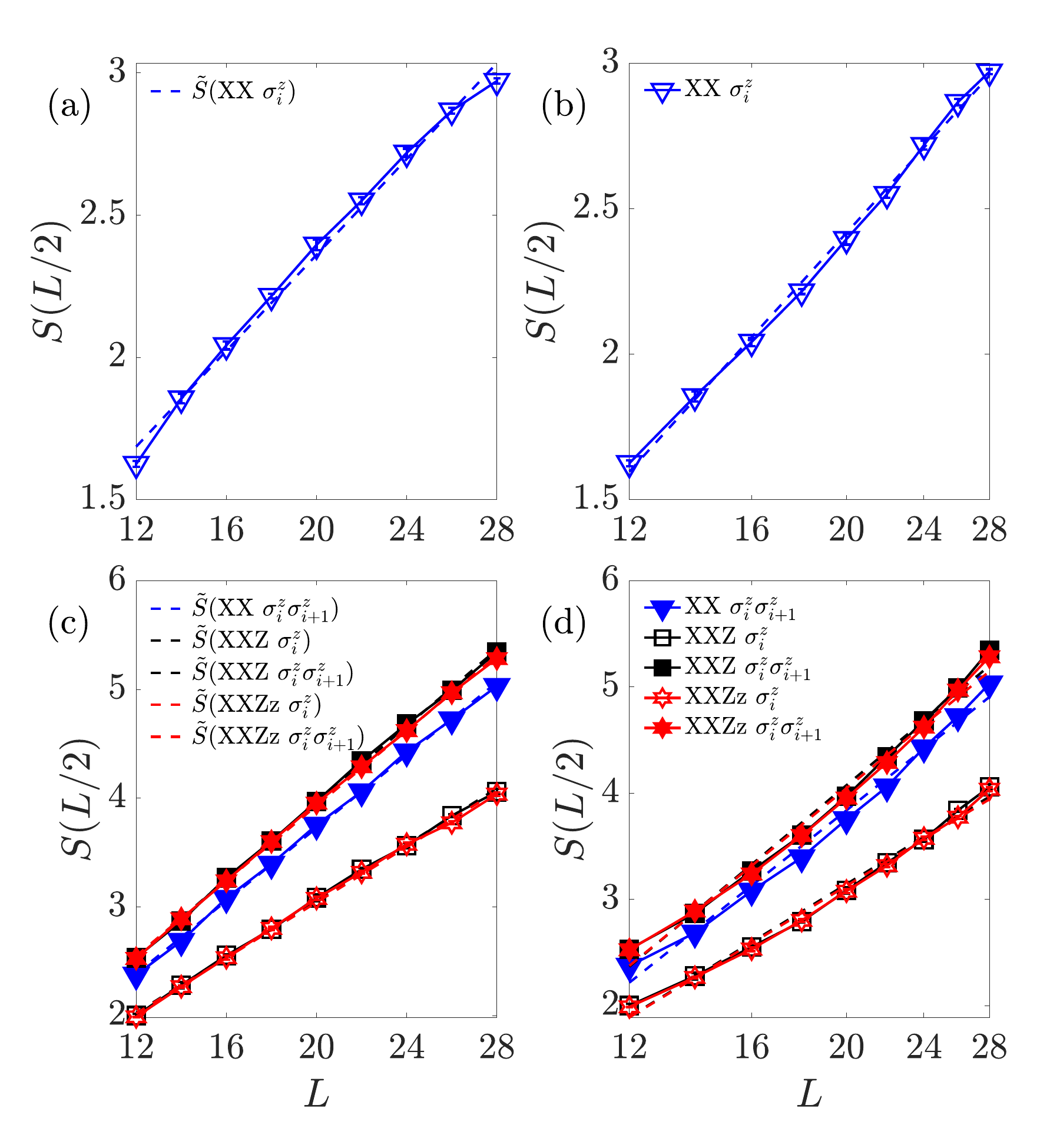}
    \caption{(a, c) The entanglement entropy $S(L/2)$ (solid line) and the linear best-fit curve in $L$, $\tilde{S}_{L}$ (dashed line), versus $L$ in linear-linear scale.
    (b, d) $S(L/2)$ (solid line) and the linear best-fit curve in $\ln(L)$, $\tilde{S}_{\ln(L))}$ (dashed line), versus $L$ in log-linear scale.
    (a, b) feature the non-interacting model and (c,d) feature the interacting models.
    All models shown in this figure preserve the U(1) symmetry.}
    \label{fig:XX}
\end{figure}

\textit{Results.---} This section studies the effect of integrability. To this aim, we consider different interaction terms, and also the role of U(1) symmetry.
For each combination of model and measurement operator, we measure the steady-state entanglement entropy $S(L/2)$ for $12 \leq L \leq 28$. Being interested in the weakly monitored phase, we fix $\gamma=0.1$ in all our simulations. 
Our main results summarizing the key aspect of our work are collated in Fig.~\ref{fig:F-test}.
In Fig.~\ref{fig:F-test}(a), we show the $F$-test $P$-values for all the models studied.
The regions close to $P \rightarrow 0$ and $P \rightarrow 1$ are enlarged in Fig.~\ref{fig:F-test}(b, c) respectively.
We see clearly that the three setups with $P \rightarrow 0$ are all and only the non-interacting ones, suggesting that they obey a sub-extensive scaling behavior.
All interacting setups, regardless of their integrability or $U(1)$ symmetry, are found close to $\mathcal{P} \rightarrow 1$, suggesting that they obey volume law scaling.
In the absence of interaction, we can expect to see a transition from volume law to non-volume law at finite $\gamma$. 
We attribute the fact that some of the models are less close to the extreme points 0 or 1 to finite-size effects.  

\begin{figure}[ht]
    \includegraphics[width=\columnwidth]{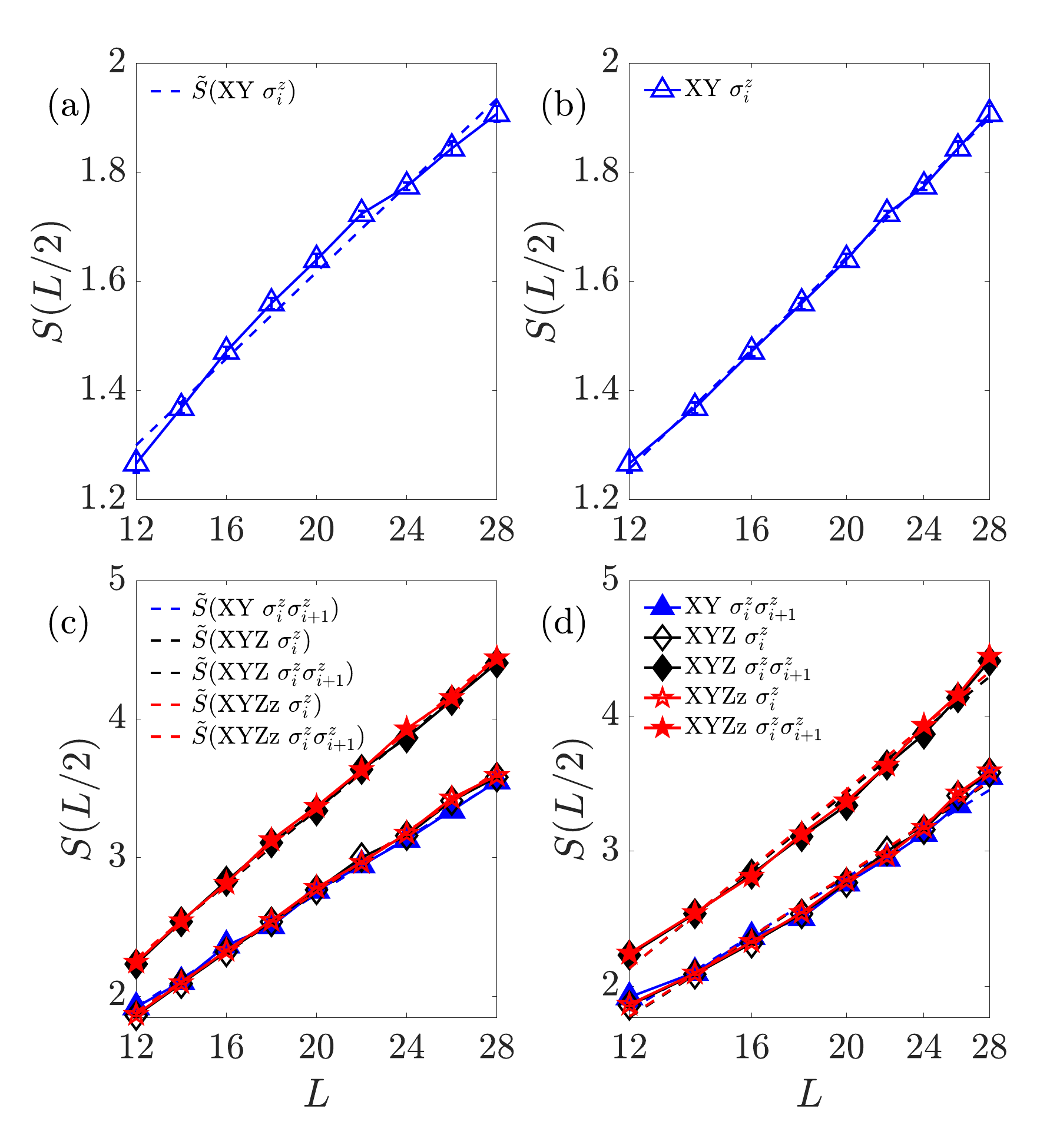}
    \caption{(a, c) The entanglement entropy $S(L/2)$ (solid line) and the linear best-fit curve in $L$, $\tilde{S}_{L}$ (dashed line), versus $L$ in linear-linear scale.
    (b, d) $S(L/2)$ (solid line) and the linear best-fit curve in $\ln(L)$, $\tilde{S}_{\ln(L))}$ (dashed line), versus $L$ in log-linear scale.
    (a, b) feature the non-interacting setup and (c,d) feature the interacting setups.
    All setups shown in this figure break the U(1) symmetry.}
    \label{fig:XY}
\end{figure}

In the rest of this section, we show the details of $S(L/2)$ scaling for various setups.
For most data points, the error bar is invisible because it is smaller than the symbols used. 
We plot $S(L/2)$ of the XX, XXZ, and XXZz models with two different measurement operators, $\hat{\sigma}_{i}^{z}$ and $\hat{\sigma}_{i}^{z}\hat{\sigma}_{i+1}^{z}$ in Fig.~\ref{fig:XX}.
These setups have one notable similarity: they preserve the U(1) symmetry.
In Fig.~\ref{fig:XX}(a, c) we plot both the steady-state entanglement entropy $S(L/2)$ (solid line) and $\tilde{S}_{L}$ (dashed line) against $L$ in linear-linear scale.
In (b, d), we plot $S(L/2)$ (solid line) and $\tilde{S}_{\ln(L)}$ (dashed line) against $L$ in log-linear scale.
We see that the only setup that deviates from the $\tilde{S}_{L}$ but follows the $\tilde{S}_{\ln(L)}$ is the XX model with $\hat{\sigma}_{i}^{z}$ monitoring, which is also the only non-interacting setup in this figure. Conversely, all interacting setups appear to grow linearly in $L$ and faster than linear in $\ln(L)$, regardless of their integrability.

\begin{figure}[ht]
    \includegraphics[width=\columnwidth]{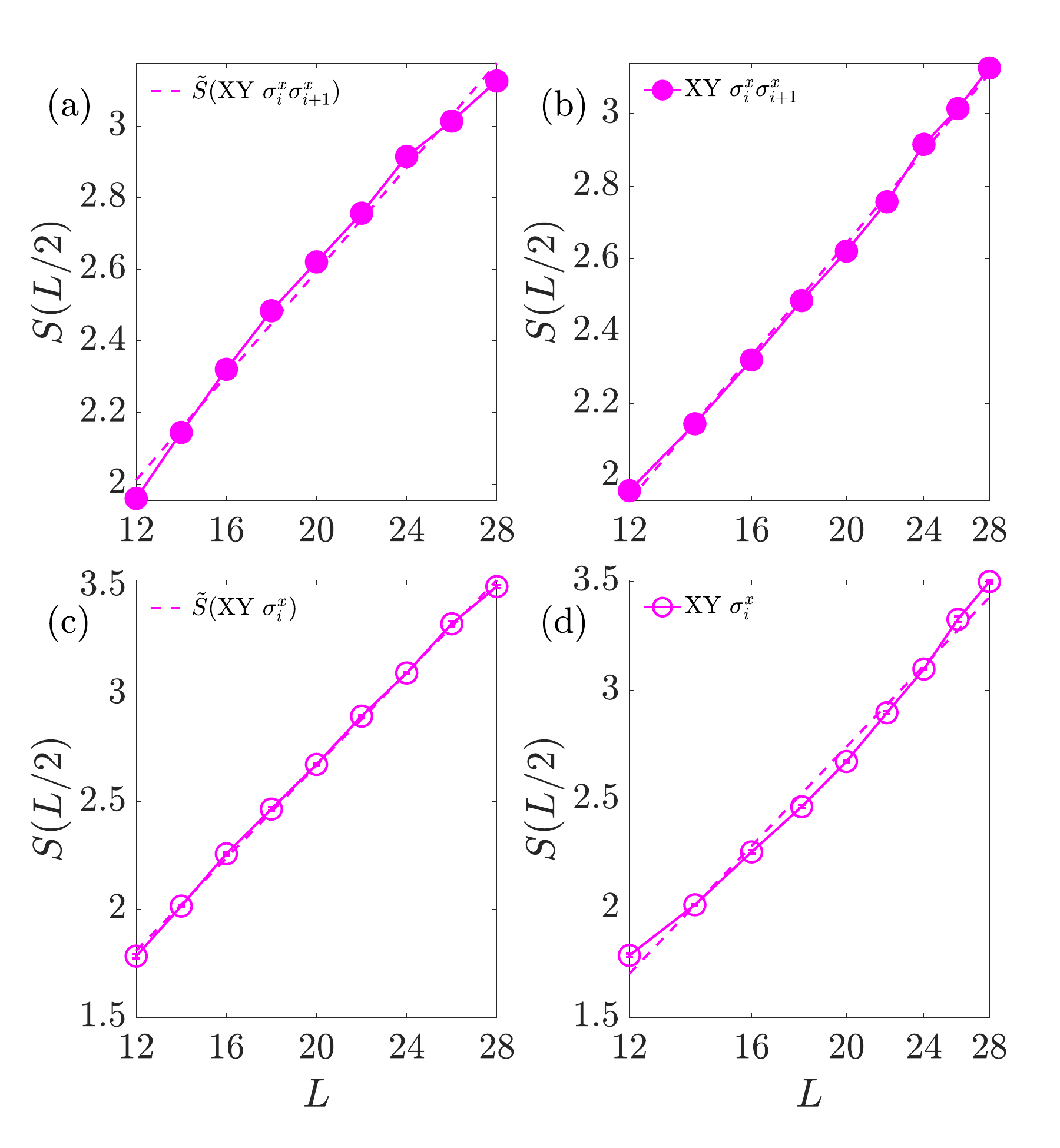}
    \caption{(a, c) The entanglement entropy $S(L/2)$ (solid line) and the linear best-fit curve in $L$, $\tilde{S}_{L}$ (dashed line), versus $L$ in linear-linear scale.
    (b, d) $S(L/2)$ (solid line) and the linear best-fit curve in $\ln(L)$, $\tilde{S}_{\ln(L))}$ (dashed line), versus $L$ in log-linear scale.
    (a, b) feature the non-interacting setup and (c,d) feature the interacting setups.
    All setups shown in this figure break the U(1) symmetry.}
    \label{fig:xXY}
\end{figure}

Next, we look at setups that break the U(1) symmetry, namely the  XY, XYZ, and XYZz models with $\sigma_{i}^z$ and $\sigma_{i}^z\sigma_{i+1}^z$ measurements.
In Fig.~\ref{fig:XY}(a, c) we plot $S(L/2)$ (solid line) and $\tilde{S}_{L}$ (dashed line) against $L$ in linear-linear scale.
In (b, d), we plot $S(L/2)$ (solid line) and $\tilde{S}_{\ln(L)}$ (dashed line) against $L$ in log-linear scale.
The only setup that deviates from $\tilde{S}_{L}$ but follows $\tilde{S}_{\ln(L)}$ is the non-interacting setup XY model with $\hat{\sigma}_{i}^{z}$ measurements.
Combining these findings, we conclude that the presence or absence of the volume law is \emph{not affected} by the U(1) symmetry, as $S(L/2)$ is not qualitatively affected by the total magnetization conservation. 

Lastly, we consider a slightly different but equally interesting scenario.
In Fig.~\ref{fig:xXY}, we study the XY model with measurements on $\hat{\sigma}^{x}$ and $\hat{\sigma}^{x}_{i}\hat{\sigma}^{x}_{i+1}$.
As previously highlighted, the XY model becomes interacting with $\hat{\sigma}_{i}^{x}$ measurements, but not when monitoring $\hat{\sigma}_{i}^{x}\hat{\sigma}_{i+1}^{x}$. 
In Fig.~\ref{fig:xXY} we see that the non-interacting XY model with $\hat{\sigma}_{i}^{x}\hat{\sigma}_{i+1}^{x}$ probes deviates from $\tilde{S}_{L}$ while following $\tilde{S}_{\ln(L)}$. Conversely, the interacting case, XY with $\sigma_{\sigma_{i}^{x}}$ probes, follows $\tilde{S}_{L}$ and deviates from $\tilde{S}_{\ln(L)}$.
Again, volume law scaling is ensured by the sole presence of interaction in the system and measurement.

\textit{Conclusions.---} 
We have studied a class of prototypical local spin chains to investigate the role of interactions, U(1) conservation, and integrability in affecting the character of the weakly measured Hamiltonian phases. In our study we focused on one-dimensional models considering systems up to $28$ spins, limiting our focus on the weakly monitored phase. 
Our differentiation of the type of measurement-induce phases is based on the $F$-test and the corresponding $P$-values. 

Our findings demonstrate that the scaling is extensive (sub-extensive) depending solely on the presence (absence) of interactions. Whereas the interactions are integrable or not, and if a U(1) conservation is present or not, does not affect the entanglement entropy scaling in the volume law phase. 
Instead, as pointed out by recent works on monitored free fermions, symmetry plays a crucial role in determining the underlying weakly-measured phase in non-interacting systems. This aspect deserves further investigation in one and higher dimensions which we leave as an outlook for future work.

This work considered the homodyne stochastic Schr\"odinger equation. It would be interesting to explore how the quantum jump type of evolution is affected by integrability, interactions, and symmetry. In that case, certain setups we consider may be interacting or not while having the same no-click (non-Hermitian) evolution dynamics~\cite{turkeshi2023entanglementandcorrelation,biella2021manybodyquantumzeno,turkeshi2022entanglementtransitionsfrom,fuji2020measurementinducedquantum,piccitto2022entanglementtransitionsin,paviglianiti2023multipartiteentanglementin}. 
From our study, we expect that measurements can lead to highly non-trivial dynamical features compared to the post-selected no-click limit.
A more systematic study in this direction is left for future work.
Lastly, locality plays a crucial role in determining the spectral and structural properties of the Hamiltonian. It would be interesting to investigate symmetry and integrability in this type of model, where the current contributions are mostly limited to power-law decaying~\cite{sierant2022dissipativefloquetdynamics,muller2022measurementinduceddark,minato2022fateofmeasurementinduced,russomanno2023entanglement} and all-to-all interactions~\cite{poggi2023measurementinducedmmultipartite,passarelli2023postselectionfree,altland2022dynamicsofmeasured,bentsen2021measurementinducedpurification,giachetti2023elusive,zhang2022universalentanglementtransitions,jian2021syk}.\\ 

{\it Acknowledgments}: D.P. acknowledges support from the Ministry of Education Singapore, under the grant MOE-T2EP50120-0019. 
We acknowledge computational resources on the College de France IPH cluster. X.T. and M.S. were supported by the ANR grant "NonEQuMat" (ANR-19-CE47-0001). This work was supported  by PNRR MUR project PE0000023- NQSTI and  by the European Union (ERC, RAVE, 101053159). Views and opinions expressed are however those of the author(s) only and do not necessarily reflect those of the European Union or the European Research Council. Neither the European  Union nor the granting authority can be held responsible for them. 
 
\bibliographystyle{apsrev4-2}
\bibliography{references}

\end{document}